# Detection of subsurface structures with a vehicle-based atom gravity gradiometer

Xiaowei Zhang[1], Jiaqi Zhong[1,2,*], Muyan Wang[1,3], Huilin Wan[1], Hui Xiong[1], Dandan Jiang[1], Zhi Li[1], Dekai Mao[5], Bin Gao[5], Biao Tang[1,5], Xi Chen[1], Jin Wang[1,2,4,†] and Mingsheng Zhan[1,2,4,‡]

[1]*State Key Laboratory of Magnetic Resonance and Atomic and Molecular Physics, Innovation Academy for Precision Measurement Science and Technology, Chinese Academy of Sciences, Wuhan 430071, China*
[2]*Hefei National Laboratory, Hefei 230088, China*
[3]*School of Physical Sciences, University of Chinese Academy of Sciences, Beijing 100049, China*
[4]*Wuhan Institute of Quantum Technology, Wuhan 430206, China*
[5]*CAS Cold Atom Technology (Wuhan) Co., Ltd, Wuhan 430074，China*
*jqzhong@apm.ac.cn
†wangjin@apm.ac.cn
‡mszhan@apm.ac.cn

High-precision mobile gravity gradiometers are critical tools in geodesy and geophysics. Atom gravity gradiometers (AGGs) could be among the most accurate mobile gravity gradiometers but are currently constrained by the trade-off between portability and sensitivity. Here, we present a high-sensitivity mobile AGG featuring an ultra-compact sensor head with a volume of only 94 L. In the laboratory, it achieves a sensitivity of 77 E/Hz$^{1/2}$ (1 E =1×10$^{−9}$/s$^2$) and a long-term stability of better than 0.5 E. We integrated the instrument in a minivan, enabling efficient mobile field surveys with excellent maneuverability in confined spaces. Using this vehicular system, we surveyed the gravitational field over a set of subsurface structures within a small wooded area, successfully resolving their structural signatures with a signal-to-noise ratio of 57 and quantifying the water depth in a reservoir with an accuracy of ±0.23 m. Compared with previous observations using a conventional gravimeter (CG-5), the superior spatial resolution is unambiguously demonstrated. This work bridges high-precision quantum sensing and practical applications in geophysics and civil engineering.

## Introduction

Precise measurement of the gravitational field holds significant importance for diverse fields such as metrology, geology, geophysics, resource exploration and civil engineering. Atom interferometer-based instruments offer distinct advantages including high precision, low drift, no mechanical wear, and the capability for long-term continuous operation [1-9]. Among them, atom gravimeters have matured sufficiently to achieve dynamic measurements on shipborne and airborne platforms [10,11], as well as high-precision mobile measurements in ground vehicles [12], demonstrating proven utility in environmental monitoring, particularly in volcanic and mountainous terrains [12, 13].

As the gravity gradient represents the second-order derivative of the gravitational potential, it is inherently more sensitive to the boundaries of anomalous masses compared to the first-order derivative (the gravity acceleration). Consequently, gravity gradiometers can, in principle, characterize subsurface density variations with higher spatial resolution than gravimeters. However, for atom gravity gradiometers [14-16], comprising two atom interferometers (AIs), both the free-fall distance of individual interferometers and the baseline length between them critically determine the instrument's sensitivity. This makes it challenging to reconcile measurement sensitivity with portability. As a result, despite achieving high sensitivity [17] and robustness [18], existing AGGs generally suffer from excessive height, large volume, and poor portability. These characteristics increase the difficulty of their integration with transportation platforms, thereby limiting the capability of mobile surveys. Consequently, the theoretical advantage of gradiometers over gravimeters has yet to be realized.

In this article, we present a high-precision compact AGG (WAGG-V-2). Leveraging a synchronous dual-fountain configuration and all-silica vacuum chambers, the instrument achieves a measurement baseline of 50 cm and a free-evolution time of $T$=150 ms within a height constraint of <1.1 m. These advancements critically underpin its sensitivity and long-term stability, approaching the best-reported values for gravity gradiometers to date. Mounted on a minivan, the system enables mobile and efficient field surveys while exhibiting excellent maneuverability in confined spaces. Utilizing this vehicular system, we surveyed the gravitational field above a set of subsurface structures within a small wooded area. Significantly, we not only resolved a dual-component structure previously undetected by CG-5 gravimeters, but also successfully estimated the water depth within one structural component (an underground water reservoir). To our knowledge, it is the first demonstration of atom gravity gradiometry's practical advantage over conventional gravimeter surveys.

## Results

**Description of the gravity gradiometer WAGG-V-2.** The AGG utilizes two vertically separated atom interferometers (AIs) to measure the differential acceleration between the two free-falling cold atom clouds. Within each AI, the cold atom cloud is split, reflected, and recombined by a sequence of three vertically propagating Raman laser pulses ($\pi/2$-$\pi$-$\pi/2$). During this process, the phase of the Raman laser pulses is imprinted onto the phase of the atom matter waves, encoding information about the atoms' trajectories. The gravitational acceleration, $g$, deflects the spatially separated interference paths, causing the atoms accumulate different laser phases and consequently acquire different matter-wave phases. This results in a phase shift for each AI, given by

$$\Delta\varphi = \Delta\varphi_1 - 2\Delta\varphi_2 + \Delta\varphi_3 = \boldsymbol{k}_{\text{eff}} \cdot \boldsymbol{g}T^2, \tag{1}$$

where, $\phi_i$ is the phase of $i$th Raman laser pulse, $k_{\text{eff}}$ is the effective wave vector of Raman laser, and $T$ is the free-evolution time between the pulses.

For two interferometers A and B separated by $\Delta z$, the phase difference caused by the differential acceleration is

$$\Delta\varphi_{\text{diff}} = \Delta\varphi_B - \Delta\varphi_A = \boldsymbol{k}_{\text{eff}} \cdot \Delta\boldsymbol{g}T^2. \tag{2}$$

The gravity gradient can then be calculated by

$$\Gamma_{zz} = \Delta\varphi_{\text{diff}} / (\boldsymbol{k}_{\text{eff}} \cdot \Delta z T^2). \tag{3}$$

When solely interested in gravity gradient rather than gravity, we can disregard the individual phases $\Delta\varphi_A$ and $\Delta\varphi_B$, along with their common-mode noise, and directly extracting the differential phase $\Delta\varphi_{\text{diff}}$ by performing a least-squares fit to the elliptical algebraic equation [20, 21]:

$$aP_A^2 + bP_A P_B + cP_B^2 + dP_A + eP_B + f = 0. \tag{4}$$

Yielding parameters $\{a, b, c, d, e, f\}$, the differential phase is then computed as:

$$\Delta\varphi_{\text{diff}} = \cos^{-1}(-b/(2\sqrt{ac})). \tag{5}$$

Here, $P_A$=1/2(1-cos($\varphi_A$)), $P_B$=1/2(1-cos($\varphi_B$)) represent the relative populations on the excited states after interference in interferometers A and B, respectively.

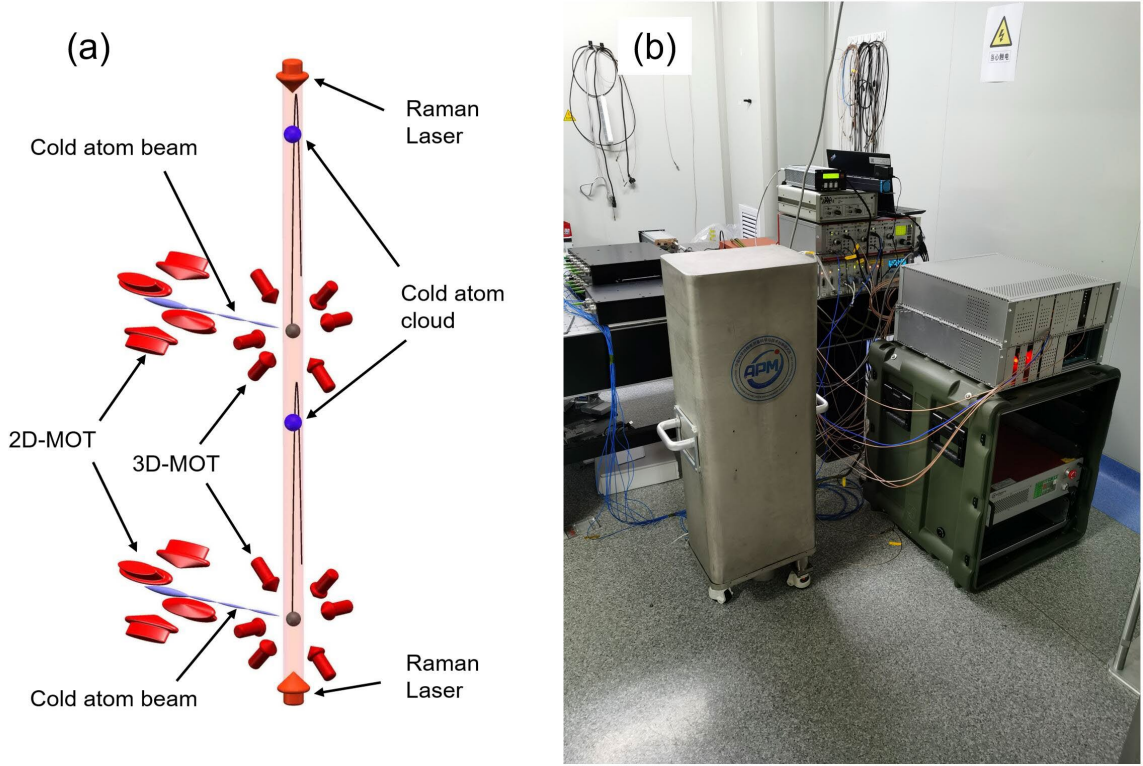

**Fig. 1** Physical schematic (a) and the photograph (b) of the gravity gradiometer WAGG-V-2. The gravity gradiometer incorporates two vertically stacked fountain-type atom interferometers. A single measurement cycle comprises sequential stages: 2D cooling, 3D cooling/trapping, atom launching, interference, and detection. For optimal space efficiency, detection occurs within the 3D-MOT zone. The complete instrument consists of three subsystems: the sensor head, the optical unit, and the electronic system. The height of the sensor head is 1.07 m.

The physical structure and the photograph of our AGG are shown in Fig. 1. During the measurement sequence, within each AI, Rb-85 atoms are first pre-cooled transversely in a two-dimensional magneto-optical trap (2D-MOT). Then they are transferred to a three-dimensional magneto-optical trap (3D-MOT) for further cooling and trapping in all three dimensions. Subsequently, the cold atom cloud is launched upward at approximately 2.0 m/s using moving optical molasses combined with polarization gradient cooling. Upon exiting the cooling region, the atoms achieve a temperature of ~3 μK. During flight, a bias magnetic field is applied to define the quantization axis. A preliminary Raman laser pulse initializes the atoms into the magnetically insensitive $|F=3, m_F=0\rangle$ state. Then the three Raman pulses ($\pi/2$-$\pi$-$\pi/2$) are applied with pulse durations of $\tau_1=\tau_3=\pi/(2\Omega_{eff})$=5 μs, and $\tau_2=\pi/\Omega_{eff}$=10 μs, where $\Omega_{eff}$ is the effective Rabi frequency. Through these light-atom interactions, the atoms acquire recoil momentum and undergo splitting, reflection, and recombination. At the end of each measurement cycle, the atom populations in the $|F=3\rangle$ and $|F=2\rangle$ hyperfine states are detected successively. The normalized atom numbers are then used for fitting to the elliptical equation (5).

The sensor head was upgraded from the V1 version documented in Ref. [19]. To maximize spatial efficiency, we retained our proprietary all-silica vacuum chambers. These chambers offer larger optical apertures than their metal counterparts, eliminate redundant inter-component connectors, and suppress eddy current effects, enabling rapid magnetic field manipulation. The chambers for the two AIs are coaxially integrated along the atomic trajectory/Raman beam axis. This configuration allows Raman lasers to propagate between atom clouds without traversing optical windows or air, thereby maximizing the performance of common-mode noise rejection. In this V2 iteration, key dimensional modifications include increasing the fountain height to 20 cm and extending the baseline to 50 cm. This change enables AI with a free-evolution time of $T$=150 ms, consequently reducing the scale factor from 8.28 E/mrad to 5.52 E/mrad. Additionally, the close proximity between 2D-MOT and 3D-MOT in the previous design caused mutual magnetic interference. To resolve this issue, we isolated the 3D-MOT and interference regions with separate magnetic shields. The final dimensions of the sensor head are 35 cm (L) × 25 cm (W) × 107 cm (H), resulting in a volume of 94 L and a weight of 66 kg. This compact design achieves "two-person portability".

In our AGG, the effective wave vector of the Raman laser is reversed between successive measurement cycles to suppress drift from non-inertial effects, thereby improving long-term stability. However, since +k and -k Raman lights impart opposite recoil velocities of approximately ±12 mm/s to the atoms, for a free-evolution time of $T$=150 ms, the atom clouds from +k and -k interferometers return to the detection zone with a position difference of up to 3.6 mm, corresponding to an arrival-time difference of approximately 1.8 ms. Because our vertically propagating detection beam is coaxial with the atomic trajectory, the positional difference of the atom cloud between +k and -k interferometers when detection is triggered results in different fluorescence collection efficiencies. To address this issue, we implemented distinct detection timing sequences for +k and -k interferometers, ensuring each cloud is interrogated at the center of the detection zone. This optimization increased the average fluorescence signal intensity by approximately 15%.

The interference ellipse and Allan deviation measured under laboratory conditions are shown in Fig. 2. The differential phase sensitivity reaches 14 mrad/Hz$^{-1/2}$, corresponding to 77 E/Hz$^{-1/2}$ for gravity gradient measurements. After 7,000 seconds of integration, the statistical uncertainty achieves 88.3 μrad, equivalent to a gravity gradient resolution of 0.49 E. Both the sensitivity and the resolution are approaching the best-reported performance levels for gravity gradiometers to date [17].

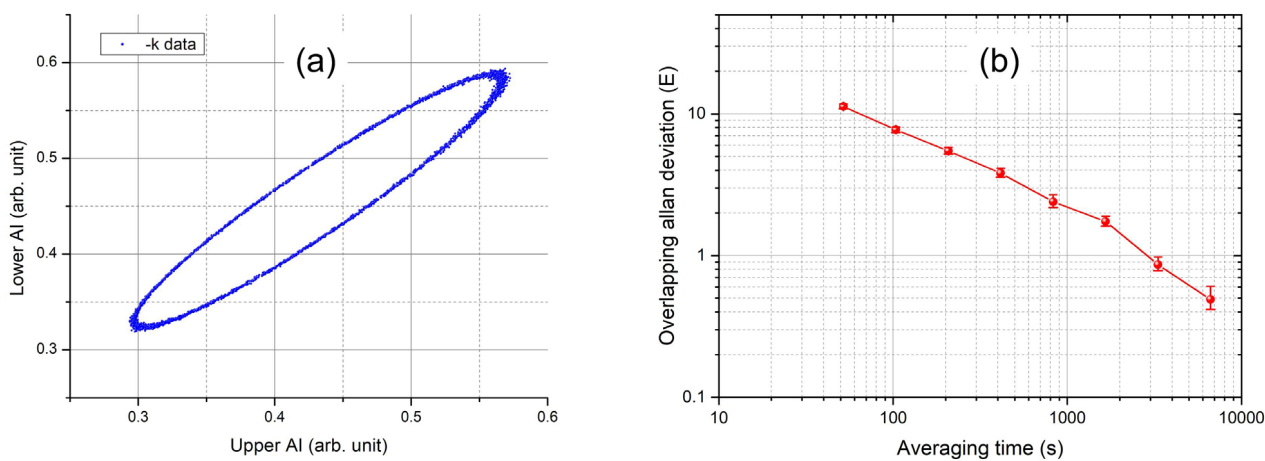

**Fig. 2** Experimental data measured by the atom gravity gradiometer (AGG). (a) Interference ellipse. The ellipse in (a) comprises 3600 data points from -k interferometry, yielding a mean elliptical phase of -318 mrad. (b) Gravity gradient Allan deviation. Each measurement phase was obtained by fitting 40 +k and 40 -k interference data points. Given a single-cycle measurement time of 650 ms, the horizontal coordinates of the Allan deviation points correspond to multiples of 52 s ($2^n$ × 52 s, n = 0, 1, 2...).

**The minivan-mounted system.** As shown in Fig. 3, our gravity gradiometer was deployed in a minivan (IVECO-FIDATO) with dimensions of 5.36 m (L) × 2.03 m (W) × 2.42 m (H). The vehicle's excellent maneuverability in narrow spaces enables surveys in areas inaccessible to larger platforms. The power supply system comprises a compact uninterruptible power supply (UPS), two high-capacity portable power stations, and a gasoline generator. The UPS remains connected to the gradiometer to provide uninterrupted power, while the two power stations alternately supply power to the UPS. The total power consumption of the gradiometer system is approximately 600 W. The UPS stores ~0.3 kWh of effective energy, capable of independently powering the instrument for ~25 minutes, which is sufficient to maintain operation during the replacement of the power stations. Each power station holds ~12 kWh of effective energy, providing ~20 hours of operation. The gasoline generator recharges depleted power stations externally, enabling infinite uninterrupted field observations. The vehicle also integrates a real-time kinematic global positioning system (RTK-GPS) for precise coordinate acquisition during large-scale surveys or in featureless terrains. Given the vehicle floor height of 0.6 m, the effective measurement height of the instrument is 1.29 m above the ground.

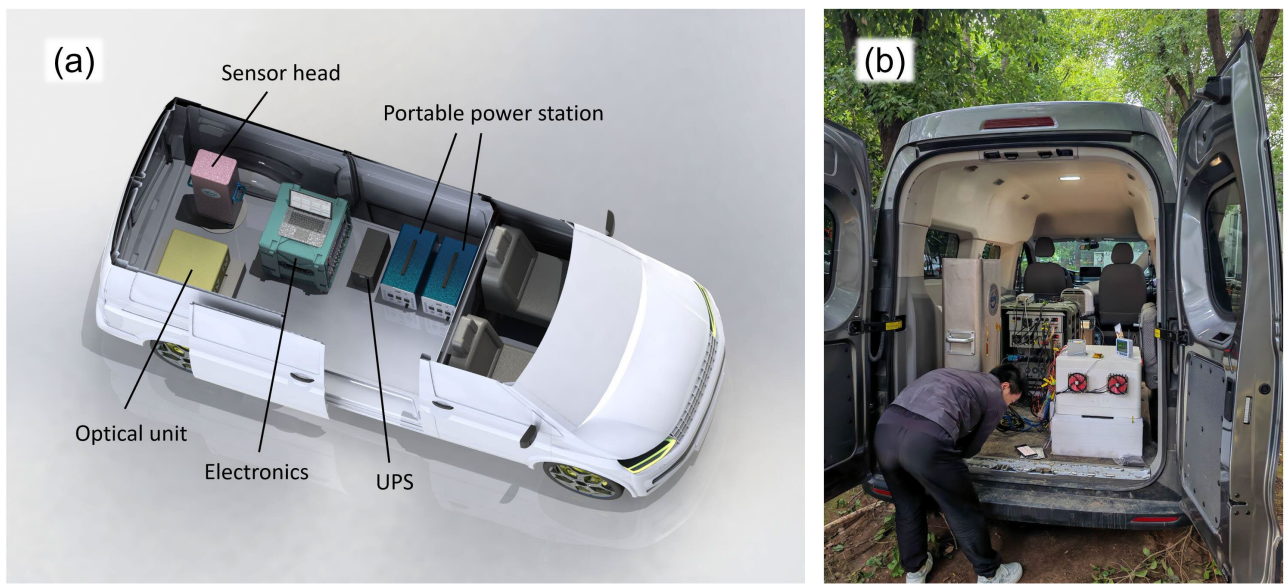

**Fig. 3** Minivan-mounted AGG system. (a) Annotated schematic of internal components. (b) Field photograph of mounted system. An operator is shown performing sensor head leveling in (b). The interior height of the minivan is 1.7 m, and at the sensor head location, there still remains sufficient space for future automatic attitude adjustment platforms or inertial stabilization platforms. The optical unit is installed in a temperature-controlled housing based on thermoelectric coolers (TECs) and placed on a shock-absorption base.

The Coriolis effect induced by Earth's rotation constitutes the dominant error source in our gravity gradiometer [19], and it has been reported as the factor that most significantly affects the repeatability during movement [18]. The elliptical phase shift in the AGG due to the Coriolis effect can be expressed as

$$\Delta\varphi_{\text{coriolis}} = 2k_{\text{eff}} \cdot (\Omega_{\text{earth}} \times \Delta v_{\text{h}})T^2, \tag{6}$$

where $\Omega_{\text{earth}}$ is the angular velocity of Earth's rotation, and $\Delta v_{\text{h}}$ is the residual horizontal velocity difference between the flying atom clouds in the upper and lower interferometers. This residual horizontal velocity difference stems from misalignments of the cooling/fountain laser paths and thus remains fixed with the instrument frame. Consequently, the resulting Coriolis phase shift varies with the instrument's orientation. The experimentally measured relationship is shown in Fig. 4. We observe that the measurement error induced by the coupling of Earth's rotation with this residual horizontal velocity difference can reach ±180.6 mrad. This error corresponds to a horizontal velocity difference of approximately 3.95 mm/s between the upper and lower interferometers. At the steepest slope of the curve, an orientation deviation of 1° can lead to a measurement error of 17.4 E—an unacceptable level of uncertainty. To mitigate this effect, we set the instrument's orientation inside the vehicle to 22.0 degrees east of north, an orientation-insensitive condition according to the curve. This confines the orientation-induced uncertainty to <3 E, avoiding the need of recalibration each time the measurement point is changed.

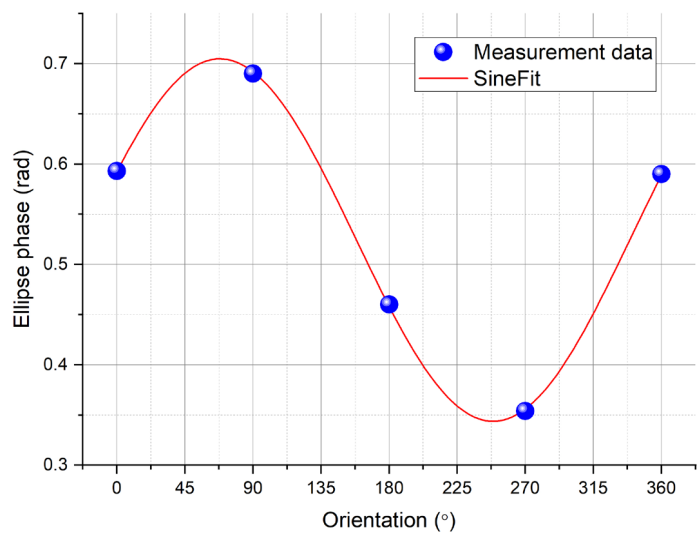

**Fig. 4** Dependence of elliptical phase on instrument azimuthal orientation. 0° corresponds to the sensor head facing east, 68.0° (22.0 ° east of north) and the 248.0° (22.0 ° west of south) are the insensitive azimuthal angles.

**Subsurface detection.** A field survey of gravity gradient was conducted in a wooded area behind Building F of the Innovation Academy for Precision Measurement Science and Technology (APM). This site was selected following our prior observation of an underground entrance and suspicion of a civil air defense tunnel. Initial investigations using a CG-5 gravimeter in the wooded area had shown a gravity anomaly of approximately 150 μGal. Leveraging the superior maneuverability of the WAGG-V-2 vehicular system, we successfully deployed the vehicle-based system within the confined woodland for gradient measurements. However, the measurement results revealed the subsurface structure to be more complex than a simple tunnel. Therefore, one week later, we coordinated with facility management to obtain access authorization, enabling detailed underground verification of the structure.

Investigations confirmed that the facility is not a tunnel but comprises two adjacent structures: a basement room and a water reservoir. As shown in Fig. 5, the room measures 9.7 m (x)×12.0 m (y)×4.9 m (z), and the reservoir measures 5.6 m (x)×12.0 m (y)×4.9 m (z), with water inside. The ground surface slopes above the structure, with the room side located at a depth of approximately 0.4 m from the surface and the reservoir side at a depth of approximately 0.6 m. Our ground survey line was positioned 1.05 m offset from the facility's x-z symmetry plane. The measurement starting point was located 4.65 m from the chamber's -x boundary.

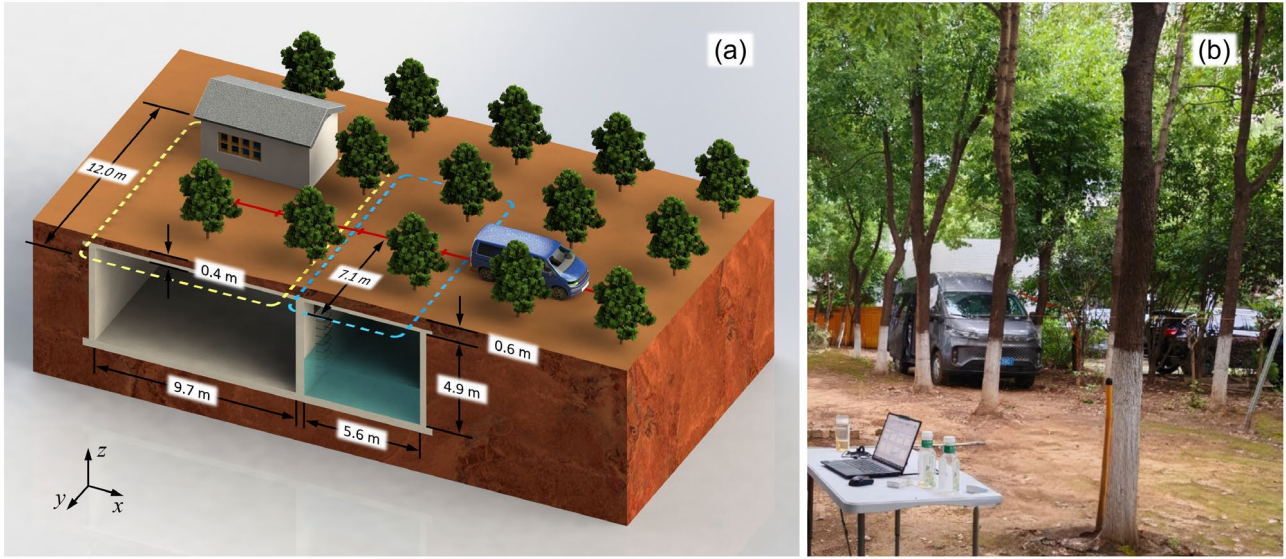

**Fig. 5** Schematic diagram (a) and photograph (b) of the test site. In (a), the yellow and blue dashed lines respectively denote the projections of the underground empty room and water reservoir onto the surface. The small building on the ground is the entrance to the underground facility. The red line and nodes on it indicate the gravity gradient survey line and measurement points.

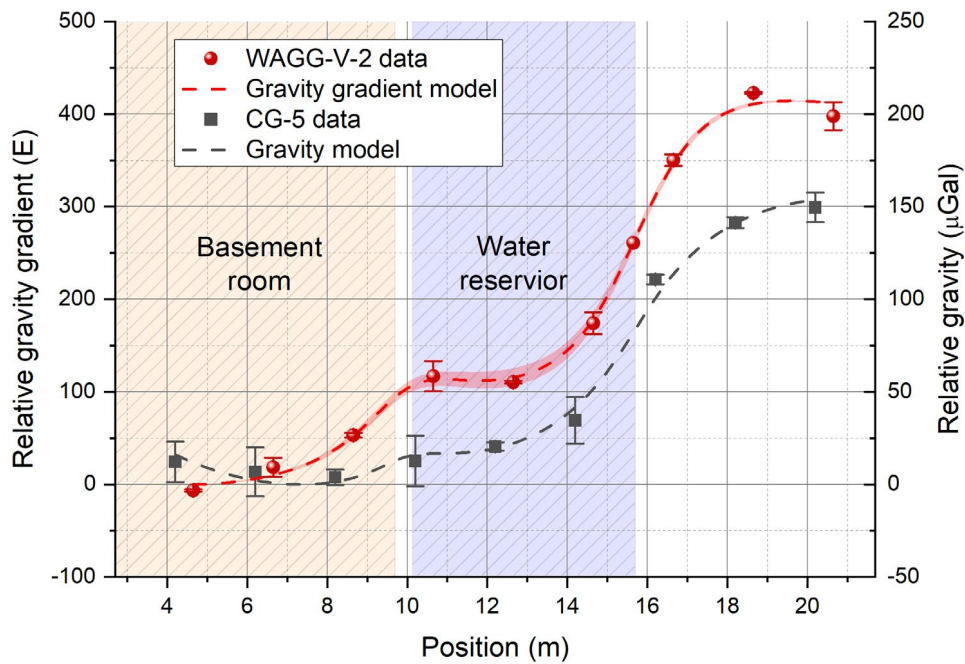

**Fig. 6** Comparison of gravity gradient measurement results (WAGG-V-2) and gravity data (CG-5 gravimeter). The pale yellow and pale blue shadings indicate the locations of the basement room and the water reservoir, respectively. The lower and upper boundaries of the red band represent the theoretical curves for water depths of 1.58 m and 2.04 m in the reservoir, while the red dashed line corresponds to the predicted depth of 1.81 m. When modeling the CG-5 gravity data, the average soil density was evaluated to ~1.83 g/cm³, approximately equal to that from the WAGG-V-2 data (~1.86 g/cm³).

Two gravity gradient surveys were conducted on May 9 and 10, 2025, respectively, along a 16 m survey line with 2 m intervals between measurement points. During the second survey, an additional point was inserted at $x$=15.5 m between the two adjacent points exhibiting significant gravity gradient changes identified in the first survey. The measurement results are presented in Fig. 6. Along the survey line, the gravity gradient exhibits a variation of approximately 414 E. At individual measurement points, discrepancies between surveys range from 2.4 E (min) to 32.5 E (max), with a mean deviation of 14.6 E. The profile reveals a distinct double-valley pattern, where the two local minima spatially coincident with the positions of the basement room and water reservoir.

We developed a theoretical model to simulate the gravity gradient generated by these subsurface structures. Based on the overall gradient variation, the average soil density in the survey area was estimated at ~1.86 g/cm$^3$ . Crucially, simulations revealed that water depth in the reservoir water depth significantly modulates the gravity gradient signature. Based on the amplitude difference between gradient minima (valleys), we estimated the water depth to be 1.81 ± 0.23 m (the red shaded area in Fig. 6). Subsequent direct measurement of the reservoir confirmed a depth of 1.85 ± 0.05 m, validating the theoretical prediction within uncertainty bounds.

Fig. 6 additionally compares these results with prior CG-5 gravimeter data. Although the gravimeter detected an overall gravity anomaly of approximately 150 μGal along the same line, it did not resolve the double-component structure with convincing signal-to-noise ratio. This contrast confirms that the gravity gradiometer offers superior sensitivity to mass boundaries compared to the gravimeter. Fundamentally, this highlights the principle-driven advantage of gravity gradiometry over gravimetry for high-resolution subsurface characterization.

## Discussion

The primary issue with this vehicular gravity gradiometer system lies in its attitude instability. Elasticity introduced by the vehicle's suspension system compromises the chassis rigidity, leading to poor wind resistance. Winds at level 3 (3.3-5.4 m/s), which slightly rocks the vehicle, significantly perturbs the atom interference signals. This frequently interrupts the measurements, thereby seriously degrading survey efficiency. Furthermore, the leveling process consumes substantial time upon reaching each new measurement point. This is exacerbated by large tilt variations between points due to uneven terrain and the vehicle's slow settlement in muddy ground. At some measurement points, completing leveling took up to 5 minutes, followed by additional waiting time for stabilization. While an expensive inertial stabilization platform would offer the most ideal solution, a viable alternative involves implementing support legs on the vehicle chassis coupled with an automatic leveling platform. This modification promises significant improvements in both wind resistance and leveling efficiency.

In summary, we designed and implemented a minivan-mounted measurement system based on a high-precision compact AGG, exhibiting excellent mobility and maneuverability in confined spaces. Deploying this system, we successfully resolved a dual-component structure under a small wooded area, previously unresolved by conventional gravimeters, and accurately predicted the water depth within the reservoir. Critically, this work demonstrates, for the first time in atom gradiometers, enhanced spatial resolution for subsurface anomalous mass detection compared to gravimeters. It thus provides a novel technological tool for applications in geodesy, archaeology, geophysical and environmental research, resource exploration, and infrastructure planning.

## Methods

**Sub-systems of the gravity gradiometer.** The optical unit adopts a design philosophy similar to Ref. [22], featuring two seed lasers and two laser power amplifiers. The seed lasers are frequency-locked to the $^{85}$Rb- $|F=3\rangle \rightarrow |F'=4\rangle$ and $|F=2\rangle \rightarrow |F'=3\rangle$ transitions, respectively, with the latter serving as the repumping laser. Seed laser for the $|F=3\rangle$ transition is amplified by the first amplifier and split into multiple parallel beams. Each beam is routed through its dedicated double-pass acousto-optic modulator (AOM) path to generate independently controlled lights, including 2D-MOT light, 3D horizontal light, fountain lights, and detection light. The Raman lasers are generated by combining the $\pm 1$ order sidebands from a 1.5 GHz AOM, followed by amplification in the second amplifier. Unlike the optical system in Ref. [22], the optical unit in this instrument employs a four-module architecture, including a seed module, beam-splitting and modulation module, and two independent power amplification modules. To address semiconductor lifespan limitations (the primary lifetime constraint), the amplification modules are designed as standalone units. This enables replacement and maintenance of these modules without disrupting other optical components. The integrated optical unit measures 36 cm × 33 cm × 21 cm (27.2 L). To adapt to large-scale temperature fluctuations in field applications, the integrated optical unit incorporates active temperature control, maintaining internal temperature stability within $\pm 0.1$ °C across ambient temperatures of 0–40 °C. Concurrently, the electronic unit was redesigned into a consolidated 12 U chassis, achieving a 50% volume reduction compared to the 24 U V1 system [19].

**Field workflow and data processing.** Two rounds of measurements were conducted on consecutive days along a 16-m survey line containing 10 measurement points (including a newly added point 7 at x= 15.6 m during the second round). At each point, we first took approximately 2 minutes to position the vehicle as precisely as possible, followed by 3-5 minutes for instrument leveling. Data acquisition then started after a 2-minute stabilization period and lasted for 20 minutes. Consequently, this workflow required approximately 30 minutes per measurement point under normal conditions. During the 20-minute acquisition period, the instrument performed 960 interferometer cycles, yielding 12 +k and -k ellipse pairs, resulting in 12 phase values. A single gravity gradient measurement value was calculated based on the average of these 12 phases.

In addition to the 10 measurement points, a base station point was established. Measurements were performed at this base station once before commencing the first measurement point and once after completing the last measurement point each day. These base station measurements were used to correct for instrument drift.

For data processing, the instrument drift rate was determined from the two base station measurements acquired each day. The measurements at each survey point were then linearly corrected for drift using the corresponding drift amount. This represents the standard practice to remove drift in geophysical surveys. Subsequently, an offset was applied to align the averages of the results from the two measurement rounds. Finally, the corrected and aligned results from both rounds were averaged point-by-point to produce the final measurements shown in Figure 6. The error bars represent the absolute difference between the two measurements at each point.

**Theoretical Modeling.** The theoretical value of the gravity gradient anomaly is obtained by numerically integrating the gravitational force generated by mass elements

$$\Delta \Gamma_{zz} = \frac{\Delta g_{\text{lower}} - \Delta g_{\text{upper}}}{d} = \frac{G\rho}{d}(\iiint \frac{\cos\alpha_{\text{lower}}}{(r_{\text{lower}} - r_0)^2}dxdydz - \iiint \frac{\cos\alpha_{\text{upper}}}{(r_{\text{upper}} - r_0)^2}dxdydz) , \qquad (7)$$

where $d$ denotes the vertical distance between interferometers in the gradiometer, $\rho$ represents the density of the mass body, $r_{\text{lower/upper}}$ and $r_0$ are the position vectors of the interferometer sensitive points and mass element respectively, and $\alpha$ is the angle between the vector $r_{\text{lower/upper}} - r_0$ and the vertical direction. For subsurface components, $\rho$ is taken as the relative density with respect to the soil density (with the soil density offset to zero). The total gravity gradient anomaly along the survey line is derived by summing the gravity gradient anomalies generated by all individual components. For example, under conditions of soil density at 1.86 g/cm³ and a water depth of 1.81 m in the reservoir, Fig. 7 illustrates the gravity gradient anomalies produced by the basement room, the reservoir, the ground building, and the total combined anomaly.

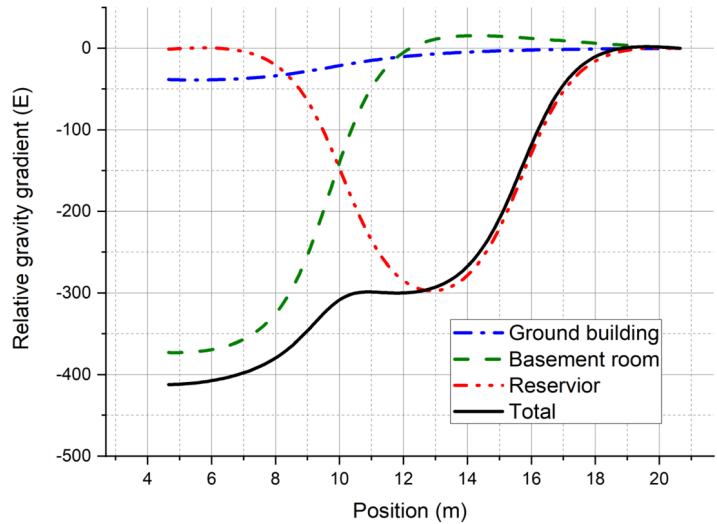

**Fig 7** Gravity gradient anomalies generated by individual components and the total combined anomaly. The anomaly from each component is also derived by summing contributions from its sub-components (e.g., the underground reservoir comprises an overlying cavity and underlying water body; the ground building is divided into roof and main sections). For simplicity, these sub-components are not explicitly decomposed here.

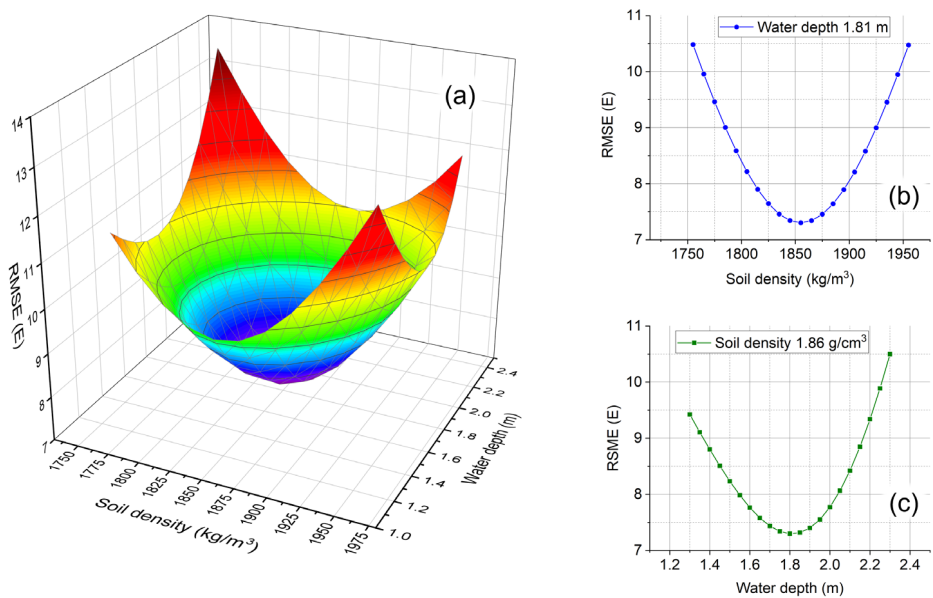

**Fig. 8** Estimation of soil density and water depth via least squares fitting. (a) Results of two-parameter scanning. (b) RMSE as a function of soil density. (c) RMSE as a function of water depth. Crucially, the two parameters

exhibit near-orthogonality, either parameter can be roughly given within a reasonable range while still allowing accurate determination of the other.

Given the predefined underground structures, the soil density and pool water depth can be estimated using the least squares method. Fig. 8 illustrates the root mean square error (RMSE) between measured and theoretical values as a function of these two parameters under the theoretical model. The minimum RMSE obtained was 7.3 E, yielding estimated values with uncertainties of 1.86 $\pm$ 0.07 kg/m³ for soil density and 1.81 $\pm$ 0.23 m for the water depth in the reservoir. In Figure 6, the theoretical gravity gradient curves (upper and lower bounds of the shaded area) corresponding to water depth variations of $\pm 0.23$ m are shown. Within the spatial range above the reservoir, the average deviation of the two curves from that for the optimal depth of 1.81 m was approximately $\pm$ 7.3 E. Relative to the overall gravity gradient signal amplitude of 414 E, the signal-to-noise ratio for this measurement is determined as 57.

**Data availability.** The datasets generated and/or analysed during the current study has been uploaded to figshare.

**Code availability.** The code that supports the findings of this study has been uploaded to figshare.

## Acknowledgements

We acknowledge the financial support from the Innovation Program for Quantum Science and Technology under Grant No. 2021ZD0300603, the Defense Industrial Technology Development Program under Grant No. JCKY2022130C012, National Natural Science Foundation of China under Grant No.12304569, and the Space Application System of China Manned Space Program (Second batch of the Scientific Experiment Project) under Grant No. JC2-0576.

We also sincerely acknowledge former team members H. Song, Y. Wang, L. Zhu, W. Lyu, D. Li, Z. Huang, W. Liu, and W. Xu for establishing foundational technologies prior to this work

## Author contributions

The design and development of the gravity gradiometer was performed by X. Zhang (the electronic system), J. Zhong (the sensor head and the optical system), and M. Wang (the automation). The optical system was built by H. Wan, with the temperature control unit from H. Xiong. The vacuum chambers were built by B. Gao. The survey measurements were contributed by X. Zhang, H. Wan, M. Wang, H. Xiong, and D. Mao. The survey site modeling and data processing were contributed by J. Zhong, X. Chen and Z. Li. The gravimeter measurements were contributed by D. Jiang and B. Tang, who also proposed this gradiometer survey for campaign. J. Zhong, J. Wang, and M. Zhan conceived and coordinated the experiment. J. Zhong, X. Zhang, J. Wang, and M. Zhan wrote the manuscript with contributions from all authors.